\documentclass[epsfig,12pt,preprint]{aastex}

\begin{document}
\title{Did SN 1989B Exhibit a Light Echo?}

\author{P.A. Milne}
\affil{Los Alamos National Laboratory, Los Alamos, N.M. 87505\\}
\author{L.A. Wells}
\affil{CFH Telescope Corp., Kamuela, HI 96743\\}


\begin{abstract}
The late light curves of SN 1989B exhibited an emission excess 
relative to other type Ia supernovae. We investigate whether this 
emission excess suggests light echo emission from this supernova, 
or rather intense background galaxy contamination at the location of 
the supernova. We conclude that stellar emission can better explain 
the late spectra, and that the emission excess is likely due to 
background contamination.
\end{abstract}

\section{Introduction}

Two type Ia supernovae (SN 1991T and 1998bu) have been determined to have 
exhibited the ``light echo" phenomenon, where dust in the vicinity of the 
SN scatters intrinsic SN light, making the SN detectable for years in 
reflected light (Schmidt et al. 1994, Cappellaro et al. 2001). 
Light echoes were suspected to be present when these supernovae (SNe) 
remained brighter at late epochs than expected from intrinsic 
luminosity sources. In both cases, light echo emission was confirmed 
when spectra taken at late times matched the continuous spectra of the 
peak emission (folded through the wavelength dependence of dust 
scattering), rather than the nebular emission expected at those epochs. 
Light echo emission from SN 1991T was further investigated with 
Hubble Space Telescope (HST) images 
of the light echo, showing both the extended nature of the echo and the 
evolution of the echo image with time (Sparks et al. 1999). 
Similarly, HST images of SN 1998bu revealed two light echo rings 
(Garnavich et al. 2001). 

SN 1989B occurred in NGC 3627 and was well-observed from 
discovery\footnote{There were multiple discoverers, including,  
Waagen, E., Knight, S., Evans, R.O., Villi, M., Cortini, G., and Johnson, W. 
(IAUC 5239).} until 
$\sim$370 days post-peak by Barbon et al. (1990) and Wells et al. (1994). 
In a 1995 survey of recent SNe designed to produce a list of light 
echo candidates for further study, Boffi et al. (1999) observed NGC 3627 
and detected fuzzy, blue emission from the location of SN 1989B. In this 
paper, we study SN 1989B in detail and assert that the available 
information suggests against this being the third SN Ia with detected light echo 
emission. 

In section 2, we compare the year one photometry with other SNe Ia and 
discuss how these observations suggest an excess of emission. In section 
3, we compare the observations of SN 1989B made years after the SN 
explosion with observations of the light echo from SN 1991T and determine 
that the fading light curves are similar. In section 4, we use the 
constraints derived from the light curve studies in an analysis of 
spectra taken within the first year after the SN explosion, determining 
that the emission excess is better explained by a collection of B-A stars 
than by the presence of light echo emission. We conclude with a discussion 
of the importance that a detection of light echo emission would 
have upon the study of SNe Ia, and how the example of SN 1989B demonstrates 
the utility of obtaining late spectra of SNe.

\section{Light Curves to +351 days}

\subsection{Evidence of Excess Emission}

SN 1989B was detected before B-band maximum and the SN was well-observed by 
many observers. The most complete study was performed by Wells et al. (1994), 
which included UBVRIJHK-band photometry and spectra at many epochs. The SN 
is understood to have been a so-called normal SN Ia, exhibiting neither the 
light curve, nor spectral anomalies seen in super-luminous SNe (i.e. 1991T, 
1992bc, etc.) and sub-luminous SNe Ia (i.e. 1991bg, 1998de, 1999by, etc.). The 
color indices indicated that the SN was highly reddened, reddening was also 
suggested by the large equivalent width of Na-D and Ca-K absorption 
lines. High-resolution spectra revealed that there were at least 
two Na-D absorption features with red-shifting that placed them in NGC 3627 
(Bolte et al. 1989).

Although the early observations suggested a SN Ia typical of the normal 
sub-class, the late observations revealed anomalous late light curves. 
Milne, The and Leising (1999) (hereafter MTL99), studied the late emission 
from eight normally- or super-luminous SNe Ia, comparing V- and B-band 
photometry with simulated light curves. Of those eight SNe, only two were 
not able to be fit by the simulated light curves, SNe 1991T and 1989B. SN 1991T 
had already been revealed to have been dominated by light echo emission after 
600 days (Schmidt et al. 1994). 
This left SN 1989B as the only SN of the eight that was not explained 
by the simulated light curves. Milne, The and Leising (2001, hereafter MTL01) 
extended their earlier work by studying the B,V,R, and I band light curves 
after 65 days of a larger collection of SNe Ia. They asserted that the late light 
curves follow two distinct shapes; 1) normally- and super-luminous  SNe Ia 
follow a similar evolution which is consistent with a single 
curve.\footnote{SNe 1991T and 1998bu were used in that study.
However, only the data before the light echoes contributed were used.}
 and 2) sub-luminous SNe Ia follow an evolution 
different from normally- and super-luminous SNe Ia and possibly consistent 
with a single curve.\footnote{Recent observations of the very sub-luminous SN 1999by 
suggests that the late light curves of very sub-luminous SNe Ia might differ 
from slightly (or transitional) sub-luminous SNe Ia, such as SN 1986G 
(Garnavich et al., in preparation).} 

Shown in Figure 1 are the V-band light curves of 
SNe 1989B, 1991T, and 1998bu compared with the light curves of 16 other 
normally- or super-luminous SNe Ia. It is apparent that initially these three 
SNe evolve similarly with the other SNe, but at later times their curves 
flatten dramatically. The data has been fit with model-generated energy 
deposition rates from the SN model DD23C (H\H{o}flich et al. 1998). 
The earlier studies (MTL99, MTL01) 
investigated positron transport in SNe Ia, concluding that at late times 
positrons escape the ejecta in quantity as would be permitted by either a 
radially-combed magnetic field or by a weak magnetic field (denoted in Figure 
1 by ``R").\footnote{The simulated light curves have finite thicknesses due to 
the allowed range of ionizations in the simulations. See MTL99 for an explanation 
of this effect.} The alternative transport scenario is that a strong, turbulent 
magnetic field traps positrons {\it in-situ}, allowing for delayed energy 
deposition, but no escape. The trapping light curves (denoted by ``T") remain 
brighter than the radial curves, but are not bright enough to explain the late 
data for SNe 1989B, 1991T and 1998bu. 
Shown in Figure 2 are the same simulated energy deposition rates fit to the 
SN 1989B data after the addition of a second light source which has 
a constant luminosity. The data are consistent with a transition from 
intrinsic emission to this additional emission. 

Excess emission is also apparent in the B,R and I band data. Shown in Figure 3 
are B,V,R, and I band observations of SN 1989B compared with other 
normally- and super-luminous SNe Ia. The ``delta magnitude" format is used in 
this figure, meaning that all data and simulations are displayed as residuals 
relative to instantaneous deposition of 100\% of the positron kinetic energy. 
The fits of the simulated light curves to the data is complicated by color 
evolution from 50 -200 days (as discussed in MTL01), however estimates of the 
amount of excess can be derived from comparisons with the other SNe. 
SN 1989B is roughly one magnitude brighter than the other SNe at +351 days, with 
B-V=0.50$\pm$0.75. 
At +307 days, we derive B-V = 0.51 $\pm$ 0.62, V-R = -0.08 $\pm$ 0.61, V-I = 
0.40 $\pm $ 0.62. 
The errors are derived from both the photometry errors and the scatter of 
the other SNe Ia and are added in quadrature. 

\subsection{Potential Sources of Emission Excess}

The data shown in 
Figures 1-3 demonstrate that there was an excess of emission in the optical 
bands at the location of SN 1989B. However, the data do not require that 
the emission is from the SN, or even if it was from the SN, it does not  
isolate a light echo as the emission source. 
If the emission excess is due to a light echo, the spectrum and flux would be 
expected to be similar to the previously detected SN Ia light echoes. 
The B-V indices derived in the previous section  are redder, but consistent 
with the values of -0.13$\pm$0.21 and $\sim$-0.1 for 1991T and 1998bu, respectively 
(Schmidt et al. 1994, Cappellaro et al. 2001). 
The constant luminosity emission source added in Figure 2
 is 6.7 magnitudes fainter than the peak V magnitude of SN 1989B. 
By comparison, SNe 1991T and 1998bu were both $\sim$9 magnitudes fainter. 
If this additional emission is entirely due to a light echo, it would 
imply a light echo about seven times more efficient at scattering the peak 
light than the other two light echoes. The images of the light echoes from 
SNe 1991T and 1998bu are very irregular, presumably because of density inhomogeneities in 
the clouds of scattering dust. It seems plausible that a scattering medium that 
uniformly possesses the characteristics of the brighter limbs of the 1991T and 1998bu 
light echoes would be able to 
enhance the total efficiency by a factor of seven. We point out that MTL99 
saw hints that the light echo for SN 1991T ``turned on" after 450 days, again 
suggestive that light echo magnitudes can vary appreciably. 

There are a number of alternative explanations for the excess emission, 
both related to the SN, and related to the host galaxy background. We will 
defer discussion of the latter to section 3 (and the Boffi et al. 1999 
observations). Of alternative SN sources of emission, the production in 
SN 1989B of additional, longer-lived radioisotopes is the most plausible.
The light curves from 100 -365 days are fit in Figures 1-3 with the energy 
deposition from the gamma-ray and positron decay products of $^{56}$Co decays. 
The decays of $^{57}$Ni $\rightarrow$ $^{57}$Co  $\rightarrow$ $^{57}$Fe 
produce gamma-rays and the decays of 
$^{44}$Ti $\rightarrow$ $^{44}$Sc $\rightarrow$ $^{44}$Ca  
produce gamma-rays and positrons, 
the energy deposition from the interactions of these decay products with 
the ejecta might be suggested to explain the emission excess.  
The primary argument against this mechanism is based on the spectra (as will 
be discussed in section 4), but the yields required to explain the 
light curve are too large, which constitutes a strong secondary argument.  Using the same 
gamma-ray and positron transport algorithms used to generate the energy 
deposition rates shown in MTL99 and MTL01, we estimated the amount of each 
isotope that would have to be produced to account for the excess. In the 
lowest-yield scenario, we determined that 1.5 M$_{\odot}$ of $^{57}$Ni (and/or 
$^{57}$Co) and 
1.0 M$_{\odot}$ of $^{44}$Ti would have to be produced, yields far in 
excess of those allowed by observations.

In summary of Section 2, emission attributed to 
SN 1989B was observed to exhibit flattening in the multi-band 
photometric evolution. This flattening is consistent with the contribution 
from a constant luminosity emission source. These observations do  
not necessitate light echo emission, but they would provide insight into the 
nature of the light echo emission, if it were to be 
determined that a light echo is present. 

\section{Detection of Emission after 5 Years}

Although not continuously monitored, SN 1989B was observed on one occasion 
since the Wells et al. observations.  B,V, and R band observations were made by 
Boffi et al.  (1999) as part of a survey of potential light echo candidates. 
The Boffi et al. 1999 observations detected emission at the 
location of SN 1989B, with V = 19.93, B-V = -0.03, and V-R = 0.02
magnitudes. The SN was considered a light echo candidate as it met their three 
criteria, 1) the emission was located near enough to the reported locations of the 
(now faded) SN, 2) the emission appeared to be from a compact patch rather than 
from a point-like source or a broadly-extended patch, and 
3) the emission was bluer than expected from clusters of background stars. 
We note that the color indices are consistent with the values derived from the 
+307 and +351 day observations.  

Wells et al. 1994 pointed out that the host 
galaxy background is bright at the location of SN 1989B, it would require observations 
with better spatial resolution to determine whether the patches are due 
to a light echo, or instead to a collection of stars. Assuming the 
emission to be due to a 
light echo, extrapolation to the Boffi et al. 1999  
observations from the Wells et al. 1994 observations suggest that the light 
echo would have faded by 1.2 magnitudes in V. After accounting for the fading 
contribution from the intrinsic emission, the decline rate of the light echo 
is found to be  0.14 mag yr$^{-1}$, roughly equal to the 0.16 mag yr$^{-1}$ 
decline rate of SN 1991T (as derived from the Schmidt et al. 1994, and the 
Sparks et al. 1999 V-band observations taken 3 years apart). 

Different geometries of scattering material can lead to different decline rates, 
and the color of the light echo would change if multiple scattering is 
dominant. Rather than discuss 
implications of the exact value of the decline rate, we emphasize that the 
emission detected 6 years after the SN explosion has both the correct 
luminosity and the correct color to be a single-scatter dominated 
light echo, within the limited precision of the data. 

\section{Spectra taken during the First Year} 

It is clear from Figure 2 that the transition from dominance by intrinsic emission 
to dominance by other emission occurs between 200 -300 days 
after the SN explosion. In addition to photometry, Wells et al. 1994 
took a sequence of spectra of SN 1989B. The
epochs of these spectra ranged from -7 days to +346 days. Shown in Figure 5 are six  
spectra from SN 1989B, taken at 100 $\rightarrow$ 365 days 
since the SN explosion (again assuming an 18 day rise to peak B magnitude). 
The sequence demonstrates that from 100 -170 days the spectra transition 
from continuum emission to nebular emission (as is typical of SNe Ia during that 
epoch), but that the 300+ day spectra appear to be continuum emission. 
These three-stage continuum-nebular-continuum transitions were also seen 
in SNe 1991T and in  1998bu and were the basis of the determining that the 
late emission from each was 
from a light echo.\footnote{We point out that both SNe 1991T and 1998bu were 
observed after 600 days, at which time the nebular-continuum transition was 
complete. The 365 day spectrum of SN 1989B is mid-transition. 
To date, there are no post-transition spectra of SN 1989B.}  
If the emission excess were due to additional radioactivities, 
the late spectra would be expected to be nebular. The fact that significant 
continuum emission is present at 365 days constitutes the primary argument 
against additional radioactivities as the source of the emission. 

We study this two-stage transition by fitting a sequence of
spectra of SN 1989B. We combine spectra from other SNe at similar epochs
first with the integrated spectrum of SN 1989B reflected off of dust, and 
second with individual stellar spectra. By comparing the ability of the 
light echo/nebular emission spectra to fit the SN 1989B spectra with the 
ability of the stellar spectra/nebular emission spectra to fit the same 
spectra, we investigate whether the emission excess is more likely to be due 
to a light echo or to background contamination. 
To generate a light echo, we blended the collection of 
spectra taken during the first 133 days, 
weighing each according to the bolometric light curve (Contardo et al. 2000). 
All of the spectra used have been reddened by dust along the line of sight, so
the integrated spectrum was de-reddened assuming E(B-V) = 0.37
(Wells et al. 1994). Near-peak spectral sequences were obtained by two groups, 
Wells et al. 1994, and Barbon et al. 1990. To avoid complications from 
different analysis methods, and to afford a glimpse at the effects of 
irregular spectral sampling upon the integrated spectrum, we treated the 
two spectral sequences separately. The two integrated spectra are shown in 
Figure 6, and appear to be largely continuous with bumps at roughly 4600$\AA$, 
4950$\AA$, 5600$\AA$, 5900$\AA$, and dips at 4900$\AA$ 
and 5750$\AA$. We use the Wells et al. 1994 spectrum for fitting, both because 
it is the better sampled spectrum, and because the late spectra we fit 
were also taken by Wells et al.. The stellar spectra were obtained from an 
on-line catalog, and are of individual stellar types.\footnote{The stellar 
spectra were obtained at http://zebu.uoregon.edu/spectrar.html.} 

The 108, 110 and 133 day spectra would be expected to be essentially 
pre-light echo (based on the light curve comparisons), 
and are comprised of both continuous emission with absorption lines, 
and of emission lines. These spectra can be reasonably well-fitted 
with the spectra from other SNe Ia paired with a light echo or a stellar 
spectrum. Shown in the upper panel of Figure 7 is the 108 day spectrum 
fitted with a combination of a light echo and a nebular spectrum from SN 1994ae 
(Bowers et al. 1997). 
Shown in the lower panel is the same spectrum fitted with the combination of a 
stellar spectrum of a A5-7V star and the nebular spectrum of SN 1994ae. 
Results of fits of these combinations to the SN 1989B spectrum are shown in 
Table 1. The addition of light echo and/or stellar emission clearly 
improves the fits to the 108 day spectra, suggesting that additional emission 
is present in the spectra as early as 108 days. This emission is likely to 
be stellar, a byproduct of host galaxy light entering the slit.  Results of 
fits to the 110 and 133 day spectra are not shown or listed due to their lack of 
coverage of the important 3500$\AA$ -4500$\AA$ wavelength range. 

The 170 day spectrum (Figure 8) demonstrates that the SN emission is 
largely nebular by this epoch. This spectrum can be well-fitted with 
the nebular spectrum of SN 1987L (Ruiz -Lapuente et al. 1993). 
The addition of a light echo or stellar 
emission improves the fit slightly, although 
there are no unique spectral features that could differentiate between the 
light echo and the stellar emission. 

The 330 day spectrum is the first that is expected to be dominated 
by the additional emission. Shown in the upper panel of Figure 9 is the 330 
day spectrum fitted with a combination of a light echo and a nebular spectrum 
from SN 1996X (Salvo et al. 2001). Shown in the lower panel is the 
same spectrum fitted with the combination of a stellar spectrum of a G9-K0V 
star and the nebular spectrum of SN 1996X. In both cases, the additional emission 
contributes primarily blue-ward of the
4700$\AA$ bump and red-ward of the 5250$\AA$ bump; these wavelength ranges are 
better fitted with the stellar spectrum than with a light echo. 
The light echo/nebular combination generates a red-ward edge to the 
4700$\AA$ feature that is not present in the 1989B spectrum. 
The near-peak spectra from both Wells et al. 1994 and Barbon et al. 1990 
clearly display a peak around 4550$\AA$. It does not seem realistic that a 
light echo could be present without it producing a feature at 4550$\AA$.
The 4900$\AA$ emission feature is not reproduced in either the light 
echo or the stellar spectrum. However, the sequence
of earlier SN 1989B spectra have shown a 4900$\AA$ bump that exceeds the same
feature in comparison spectra. Thus, the 4900$\AA$ bump might be
nebular emission that is poorly modelled with the SN 1996X spectrum
(Liu et al. 1997 attributes the nebular emission at 4900$\AA$
to a pair of FeII/FeIII lines). 

A second late spectrum was obtained by Wells et al. 1994, at day 365. This 
spectrum was fitted similarly to the 330 day spectrum. The resulting fits 
are shown in Figure 10. The same nebular spectrum of SN 1996X was used for 
comparison, as the shape of the nebular spectrum 
is not expected to change very much from 
330 -365 days. It is clear that the continuous emission in this spectrum  
is quite different than the continuum seen in the 330 day spectrum. In 
addition, the H$\alpha$ line observed at 6600$\AA$ is an emission line in the 
365 day spectrum, whereas it is an absorption line in the 330 day spectrum. The 
light echo's angular diameter would be expected to be less than 6 mas at one 
year, which would be a point source compared with the 0.4 arcsec pixel$^{-1}$ 
slit width of the 
instrument. Since the nebular emission was detected in both the 330 day and 365 
day spectra, it is likely that the variation in the spectra is due to differences 
in the background sampling rather than due to differences in the sampling of 
the light echo. In the lower panel of Figure 10, the 365 day spectrum is shown 
fitted with the nebular emission from SN 1996X combined with a stellar spectrum from 
an A1-3V star. Again the stellar spectrum affords the better fit. 
As with the 330 day spectrum, the light echo generates an absorption feature 
at 6150$\AA$ that is not present in the 365 day spectrum. Also, the 4700$\AA$ 
bump is better fitted by the stellar spectrum than by the light echo. There 
are absorption features in the 365 day spectrum that suggest Balmer lines 
(as would be expected from stellar spectra of this spectral type), although the 
spectrum does not reach to short enough wavelengths to fully sample the Balmer 
discontinuity. 

In summary, spectra taken of SN 1989B between 100 -170 days after the explosion 
reveal it to have been a relatively normal SN Ia, becoming increasingly 
dominated by nebular emission lines. The addition of stellar and/or light echo 
emission as secondary components improves the fits to these spectra. Two spectra 
taken after 300 days exhibited substantial continuum emission, which is not 
expected from SN Ia at that epoch. 
The continuous nature of these spectra argue against additional radioactivities 
as an explanation for the excess emission seen in the light curves. 
The spectra can be suitably reproduced with combinations of nebular emission 
plus stellar spectra, with these combinations affording better fits overall than 
combinations of nebular emission plus light echo emission. The two late spectra exhibit 
different continua, this is likely due to differing background contamination from 
two different slit orientations. The preference for nebular/stellar emission 
combinations evident in Table 1 does not preclude a light echo contributing at some, 
lower level, which might be observable in a future observation. 
The light echo spectrum is not rich with structure, but in the absence of 
the nebular emission (as would be the case if a spectrum were obtained now) 
a light echo would be apparent in a sufficiently high S/N spectrum.

\section{Discussion} 

The declining light curves, late images, and  one-year spectrum of SN 1989B all 
appear to be consistent with the existence of an emission excess at the location 
of the SN. If a light echo from this SN is the dominant contributor of this 
emission excess, it is more efficient (by roughly a factor of seven)  
at reflecting light than the previous two SN Ia light echo detections, 
SN 1991T and SN 1998bu. However, the current collection of observations 
suggest against the dominance of light echo emission as a source of this 
emission excess, favoring instead background contamination. 
Bearing in mind that it has been 
suggested that light echoes might be detectable for decades, the opportunity 
remains to further study this region.  Two promising observations would 
be a high-resolution image of the SN 1989B region, and a high S/N spectrum of 
the region (now that the intrinsic, nebular emission has faded). The first 
observation has been selected for the HST-ACS (as a GTO observation), 
the second would be a good observation for a large ground-based telescope. 

Whether or not the emission excess is due to a light echo, the late spectra 
suggest that the emission is not intrinsic to the supernova. This removes 
SN 1989B as an exception to the tendencies claimed at late 
times for SNe Ia (i.e. that the late light curves show evidence of positron 
escape). Although it is unlikely that a light echo 
dominates the emission excess, a light echo could be present at some level. 
If present, a light echo could be used to probe the medium 
surrounding SN 1989B, and for polarization images of the light echo to be used 
to estimate the distance to NGC 3627. 

\begin{table}
\caption{Fits to SN 1989B spectra. Best-fitting spectrum in {\bf bold}.}
\begin{tabular}{cc|ccc|ccc|c}
\hline
\hline
Epoch & Wavelength & \multicolumn{3}{c|}{Neb. + Echo} & 
\multicolumn{3}{c|}{Neb. + Stellar} & Neb. -Only \\
$[d]$ & Range $[\AA]$ &  $\sigma^{a}$ & N:LE$^{b}$ & $\alpha^{c}$ & $\sigma^{a}$ &
 N:Stel.$^{d}$ & Type & $\sigma^{a}$ \\
\hline
108 & 3500-8000 & 2.77 & 3.2 & 0.1 & {\bf 2.48} & 1.5 & A5-7V & 3.57 \\
170 & 3540-6220 & 0.49 & 3.4 & 0.3 & {\bf 0.43} & 2.3 & A5-7V & 0.62 \\
330 & 3680-6700 & 0.31 & 0.7 & 0.1 & {\bf 0.20} & 0.4 & G9-K0V & 0.46 \\
365 & 3950-6700 & 0.18 & 0.4 & 0.7 & {\bf 0.14} & 0.2 & A1-3V & 0.44 \\
\hline
\end{tabular}
\begin{tabular}{cl}
$^{a}$ & Standard deviation of spectral fit in units of
10$^{-16}$ erg cm$^{-2}$ s$^{-1}$ $\AA^{-1}$ \\
$^{b}$ & Nebular to light echo ratio in the 3500$\AA$ -9700$\AA$ wavelength range. \\
$^{c}$ & Scattering index of light echo ($\lambda^{-\alpha}$). \\
$^{d}$ & Nebular to stellar emission ratio in the 3500$\AA$ -9700$\AA$
wavelength range. \\
\end{tabular}
\end{table}

\begin{figure}
\plotone{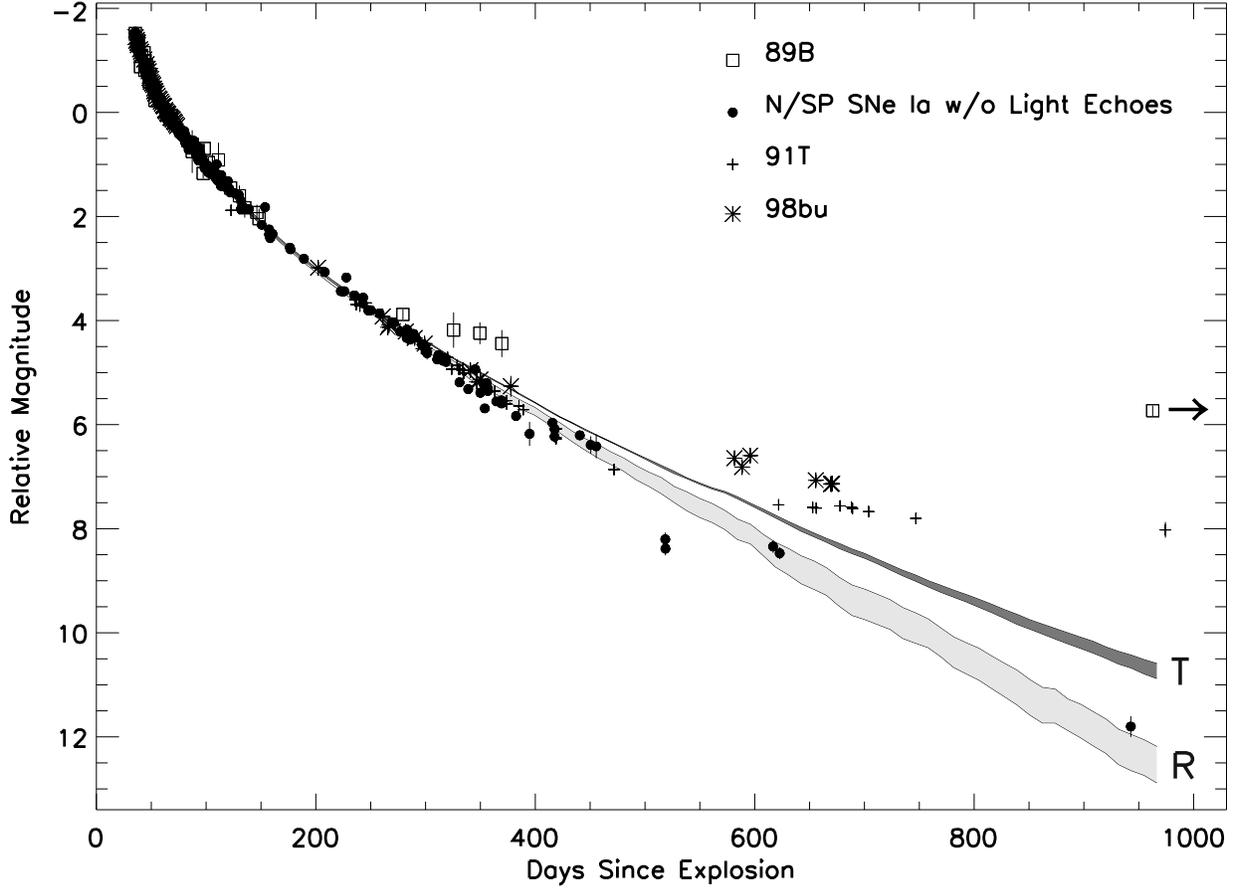}
\caption{Late light curves of Type Ia SNe. V-band photometry of SNe 1989B, 1991T, 
and 1998bu compared with 16 other SNe Ia fitted with the SN model, DD23C (H\H{o}flich 
et al. 1998). SN 1989B (open squares) and SNe 
1991T \& 1998bu (crosses) remain brighter at late times than the other SNe Ia 
(filled circles). All data and simulations are shifted to have zero magnitude 
at day 65. Simulations that permit the escape of positrons through radially 
combed magnetic field lines are shown with light shading (R). Simulations that 
trap positrons with a strong, turbulent magnetic field are shown with 
dark shading (T). Data for SN 1989B is from Wells et al. 1994, Barbon et al 
1990, and Boffi et al. 1999. 
The latest observation of SN 1989B was made at 2294 days after the explosion, 
not 960 days as plotted.
Data for SN 1991T is from Schmidt et al. 1994, Lira et al. 1998, 
Cappellaro et al. 1999. Data for SN 1998bu is from Jha et al. 1999, Suntzeff et 
al. 1999, Garnavich et al. 2000, unpublished. Other SNe used are:  
SNe 1990O,1991ag,1992al,1992bc,1993ag from Hamuy
et al. 1996.,  SNe 1995D,1995E,1995ac,1995al,1995bd,1996X from Riess et al.
1997, SN 1992A (Suntzeff 1996), 1990N (Lira 1998),
1991T (Lira 1998; Schmidt et al. 1994), SN 1994D (Patat et al. 1996, Tanvir 1997, 
Cappellaro 1997,1998). 
SN 1996X is also from Salvo et al. 2001, SN 1995D also from Sadakane et al. 1996.
\label{fig1}}
\end{figure}

\begin{figure}
\plotone{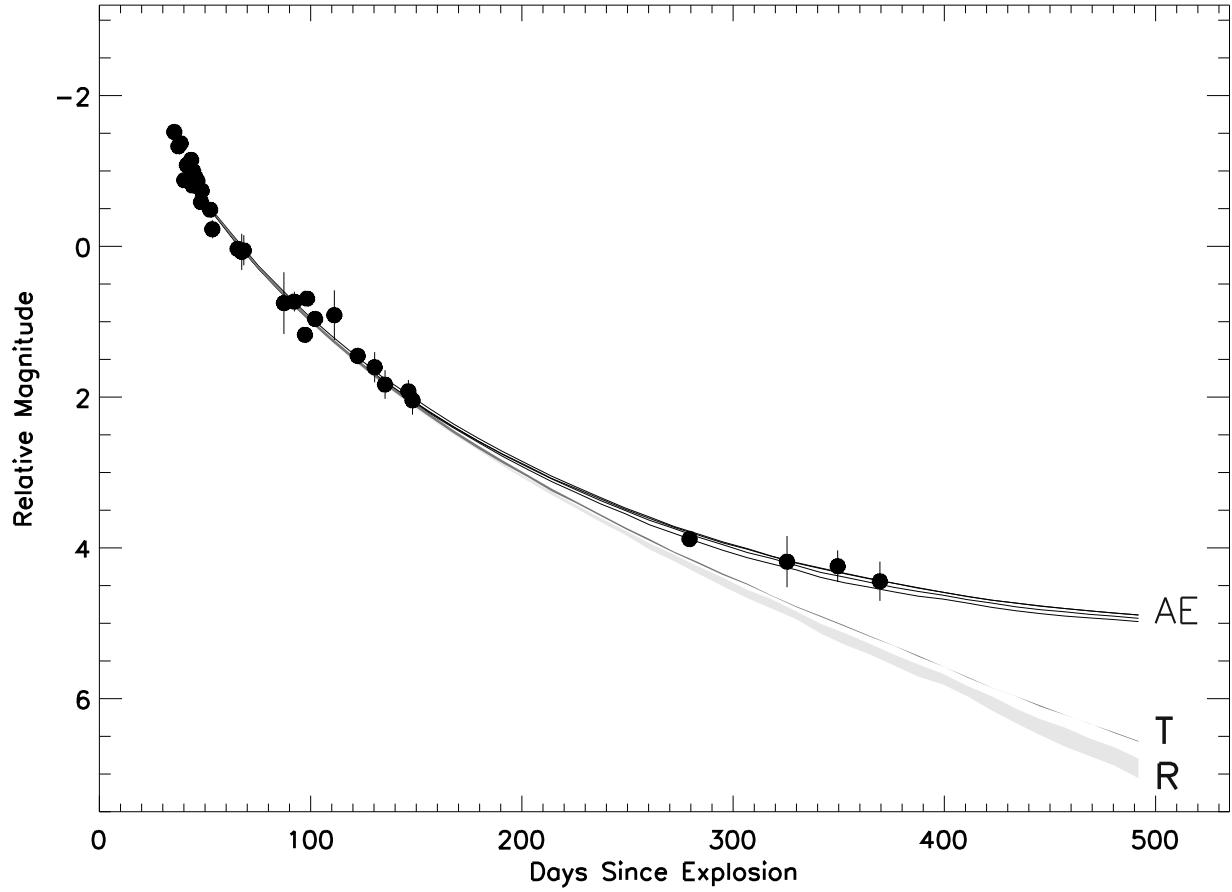}
\caption{V-band light curve of SN 1989B fitted with the energy deposition 
rates of the SN Ia model DD23C (H\H{o}flich et al. 1998). 
The model fits are shown both with and 
without the contribution of a constant luminosity light source.  Neither 
the radial curve (R), nor the trapping curve (T) 
can explain the data without the contribution of an additional emission 
source (AE). The data is from Wells et al. 1994, and Barbon et al. 1990. }
\label{lecho_89b}
\end{figure}

\begin{figure}
\plotone{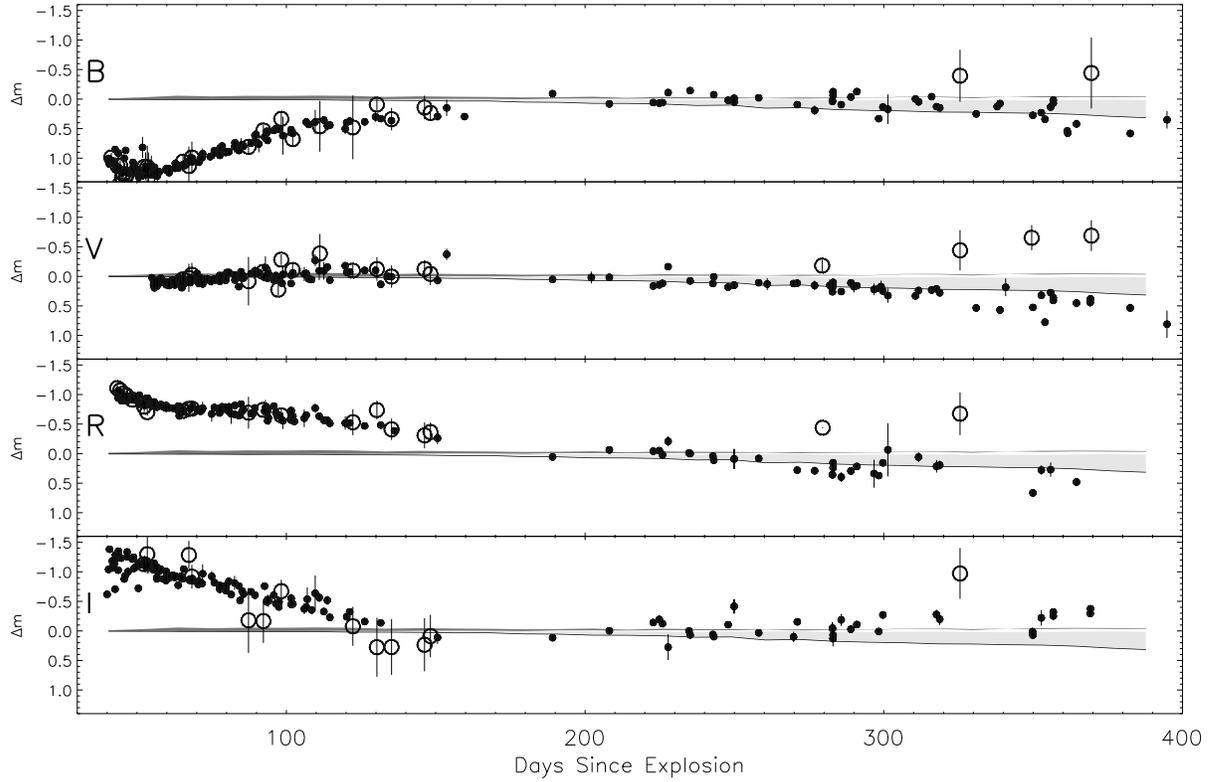}
\caption{BVRI band light curves of SN 1989B compared with light curves of 
16 other SNe Ia and with the SN Ia model, DD23C. 
SN 1989B (open circles) follows an evolution similar to the other 
SNe (filled circles) initially, but is brighter at late epochs in all four 
bands. Photometry and model references are as in Figure 1. All data is plotted on a 
magnitude scale relative to instantaneous deposition of 100\% of the positron 
kinetic energy. The trapping light curves appear as a single line in this figure 
due to the earlier epoch plotted relative to Figure 1.}
\label{nmhst}
\end{figure}

\begin{figure}
\plotone{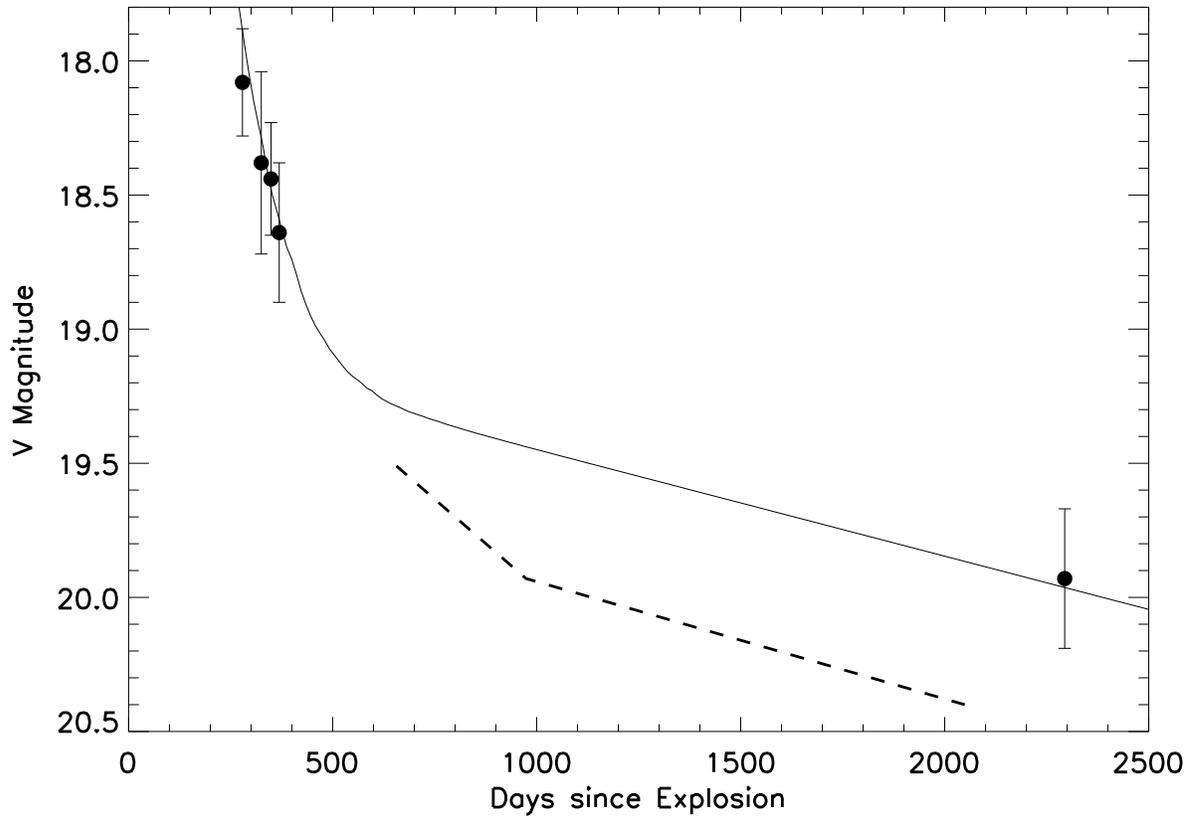}
\caption{Late observations of SN 1989B. Late V-band observations of SN 1989B 
(filled circles) are fitted with an emission excess that loses 0.145 magnitudes per year 
(solid line). For 
comparison, V-band observations of SN 1991T are shown (dashed line), after 
being shifted by -1.5 magnitudes. SN 1989B data is from Barbon et al. 1990, 
Wells et al. 1994, Boffi et al. 1999. The SN 1991T data is from Schmidt et al. 1994, 
and Sparks et al. 1999. }
\label{late89b}
\end{figure}

\begin{figure}
\plotone{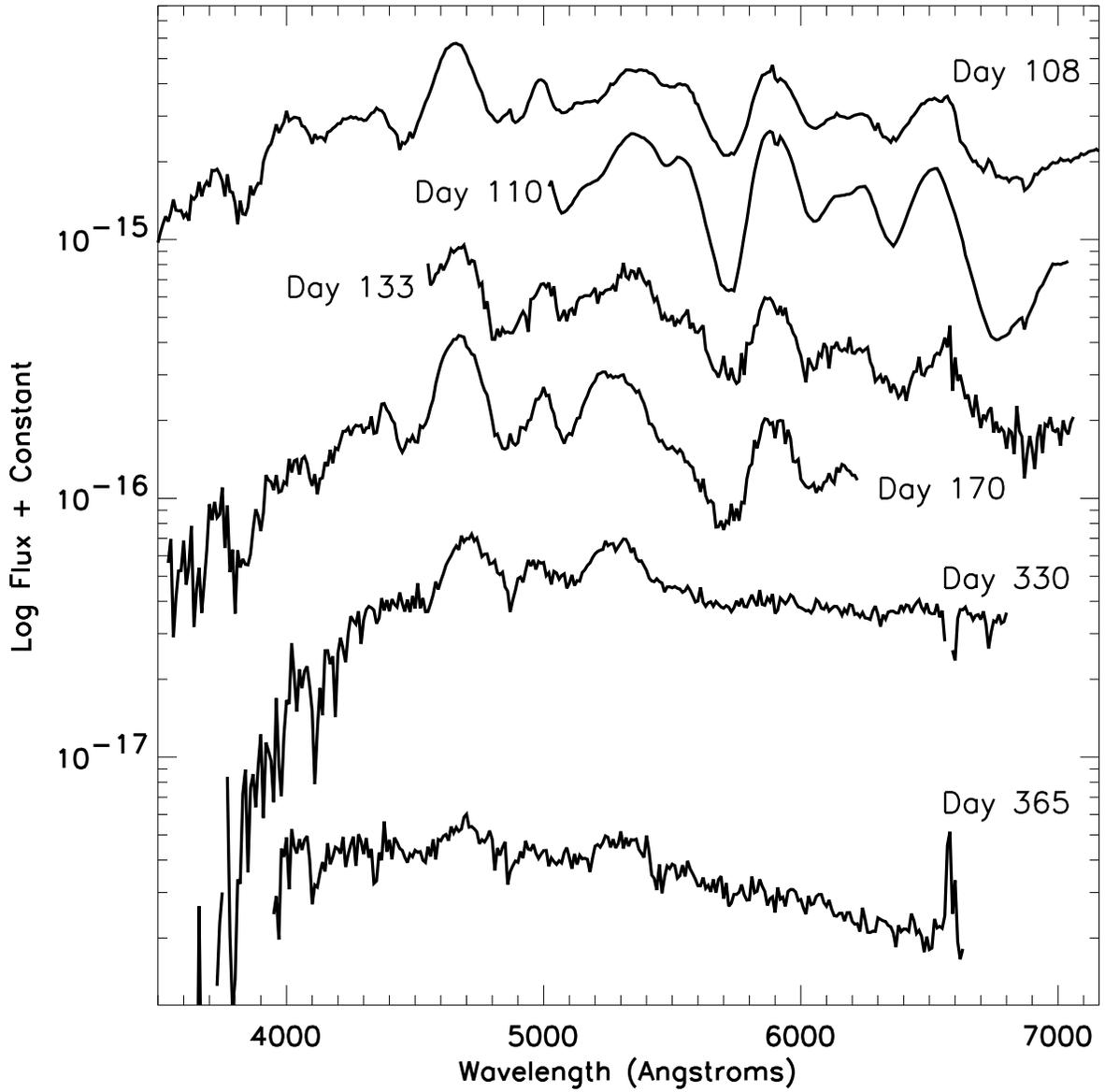}
\caption{A sequence of spectra of SN 1989B. The early and late spectra are 
clearly continuum-dominated, while the middle spectra are nebular. The late 
spectrum is unlike those observed in SNe Ia, being instead similar to the 
light echoes from SNe 1991T and 1998bu. All spectra are from Wells et al. 1994.} 
\label{spseq}
\end{figure}

\begin{figure}
\plotone{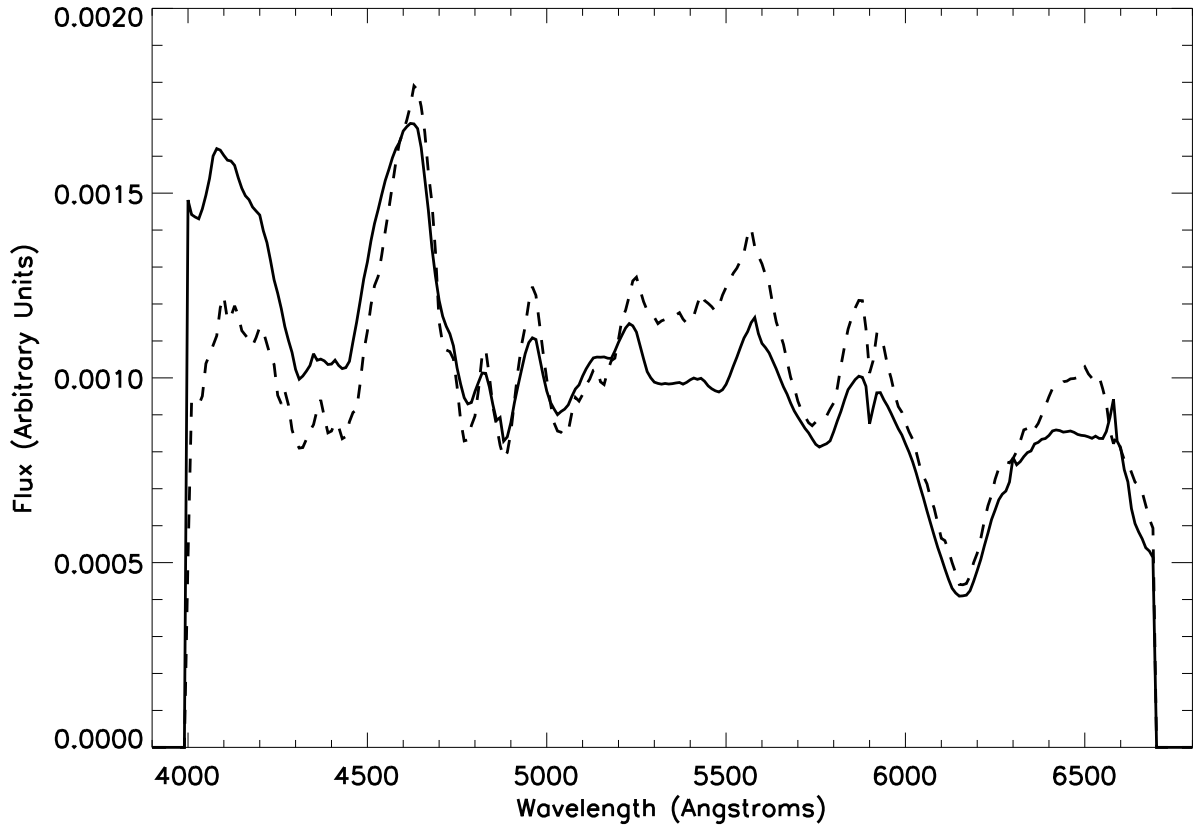}
\caption{Integrated spectra of SN 1989B. The integrated spectrum derived 
from Wells et al. 1994 spectra (solid line), agrees fairly well with the 
integrated spectrum derived from Barbon et al. 1990 spectra (dashed line).} 
\label{wb}
\end{figure}

\begin{figure}
\epsscale{0.85}
\plotone{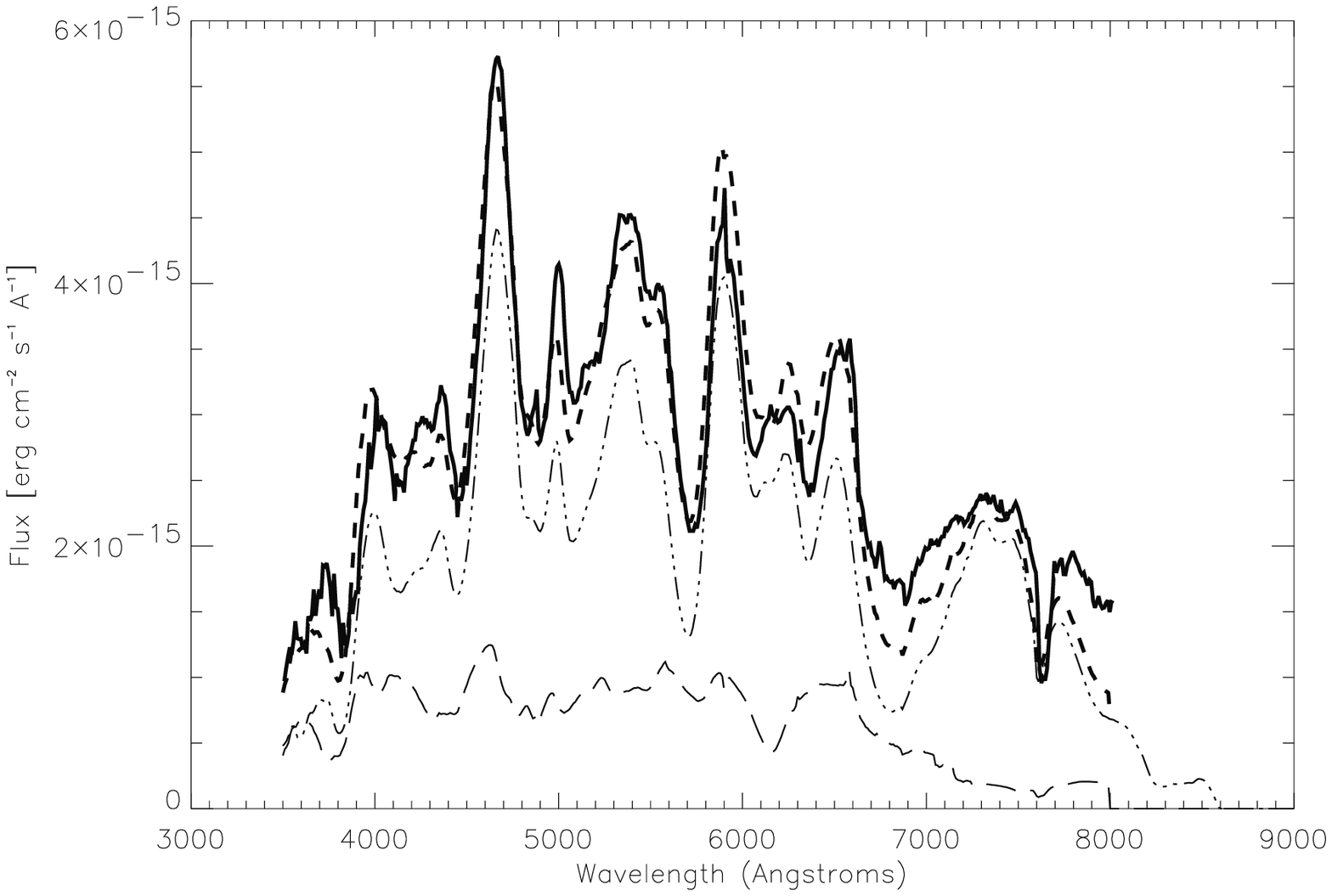}
\plotone{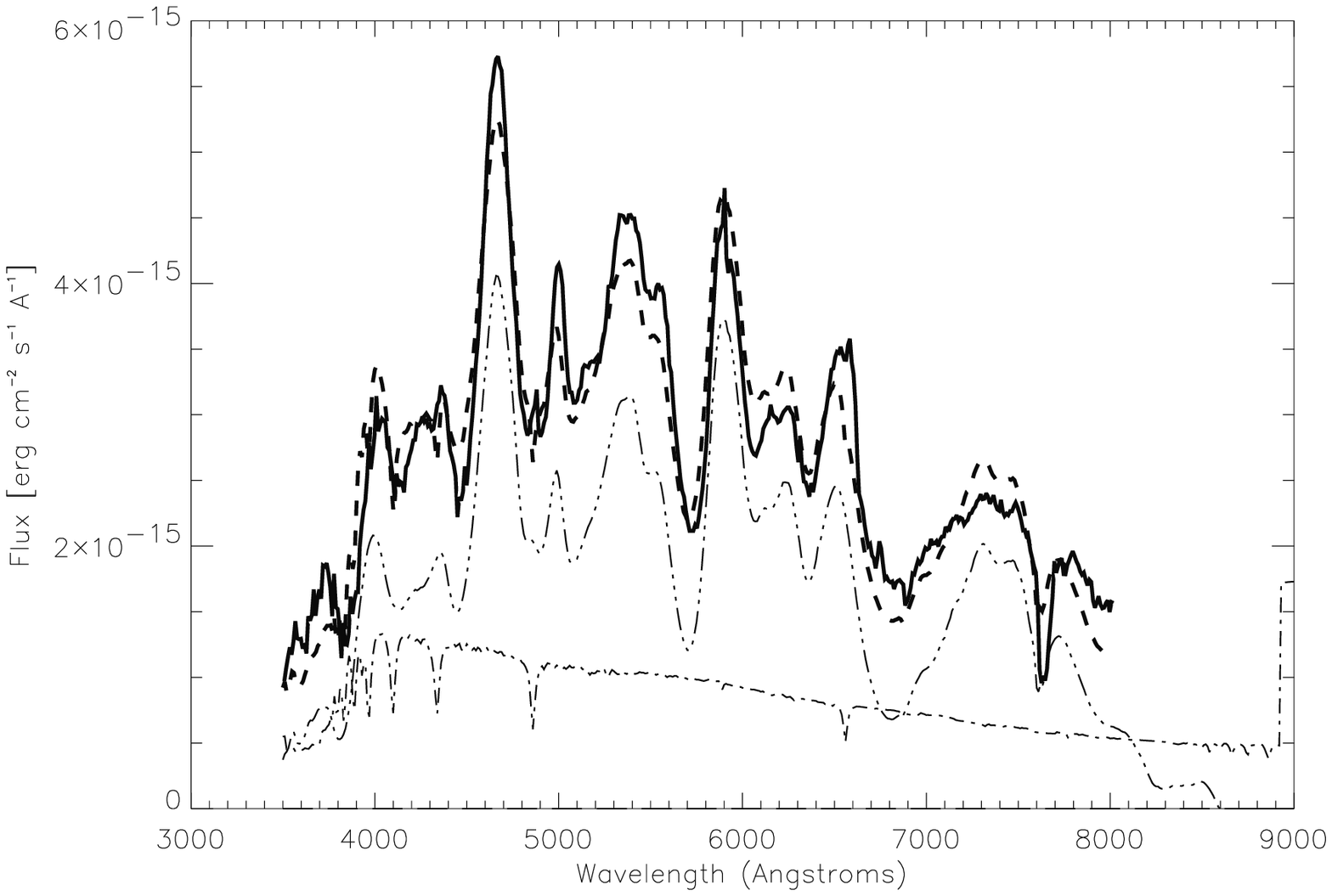}
\caption{The day 108 spectrum of SN 1989B. In the upper panel, 
the spectrum (solid line) has been fit with a day 106 spectrum of SN 1994ae 
(dot-dot-dot-dashed line) and a faint light echo (long-dashed line). 
In the lower panel, the spectrum has been fit with the SN 1994ae 
spectrum and a faint A5-7V stellar spectrum (dot-dashed line). 
In both panels, the  relative contributions to the composite spectrum  
were determined by least square fitting. The combined 
spectra (thick dashed lines) reproduce the principal features of the 1989B 
spectrum, and it is likely that the differences between the spectra are due 
to heterogeneity amongst normal SN Ia spectra. 
Reddening of E(B-V)=0.37 magnitude was assumed.}
\label{day108}
\end{figure}

\begin{figure}
\epsscale{0.90}
\plotone{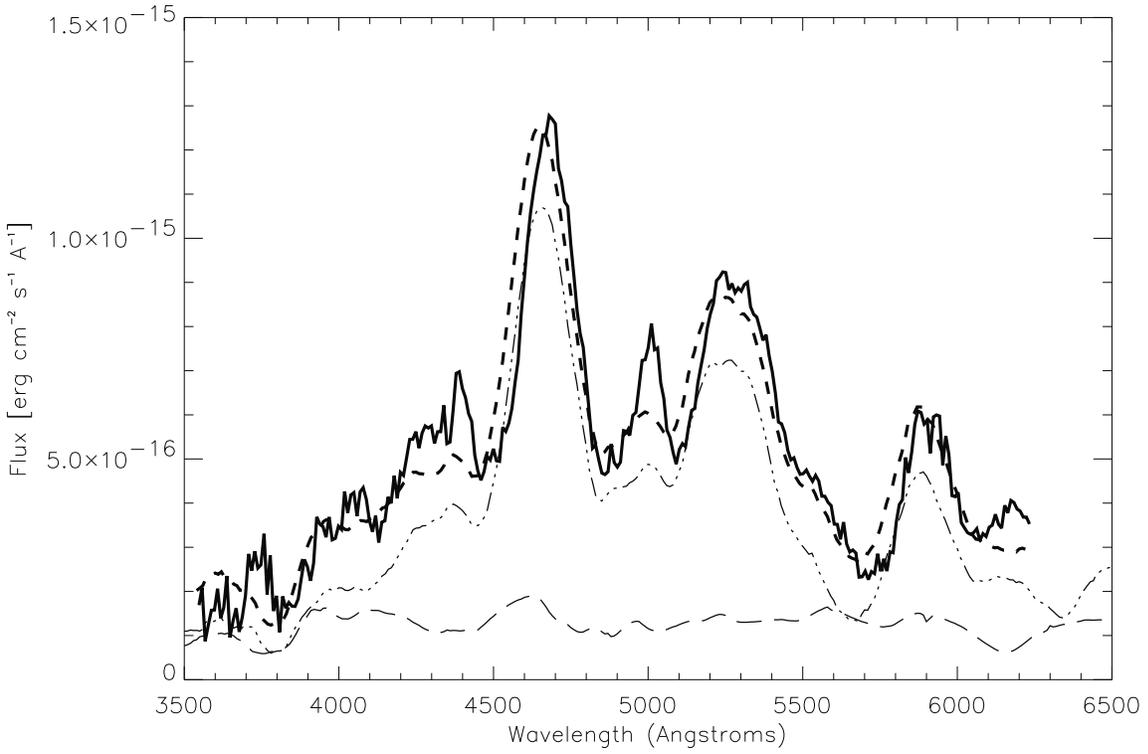}
\plotone{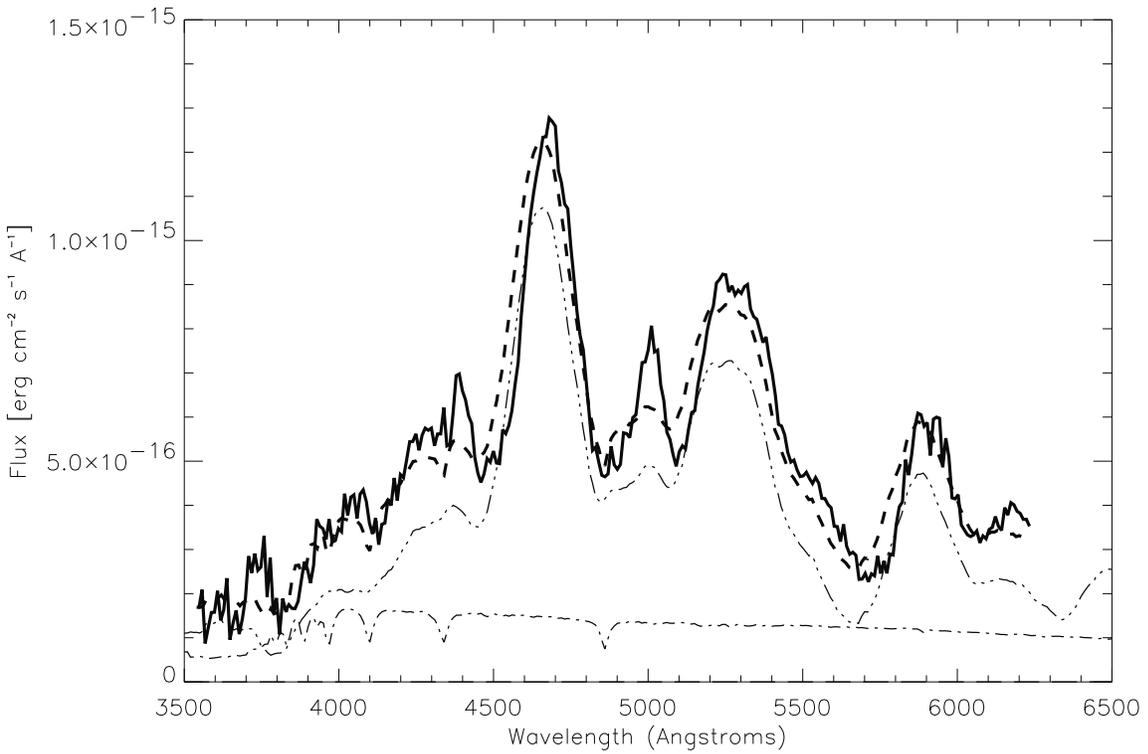}
\caption{The day 170 spectrum of SN 1989B. The spectrum is fit as in 
Figure 7. The nebular spectrum is a day 168 spectrum of SN 1987L, 
the stellar spectrum is of a A5-7V type star. 
The SN 1989B spectrum is largely nebular, and is well fitted by 
the two simulated spectra.  Reddening of E(B-V)=0.37 magnitude was assumed.}
\label{day170}
\end{figure}

\begin{figure}
\epsscale{0.90}
\plotone{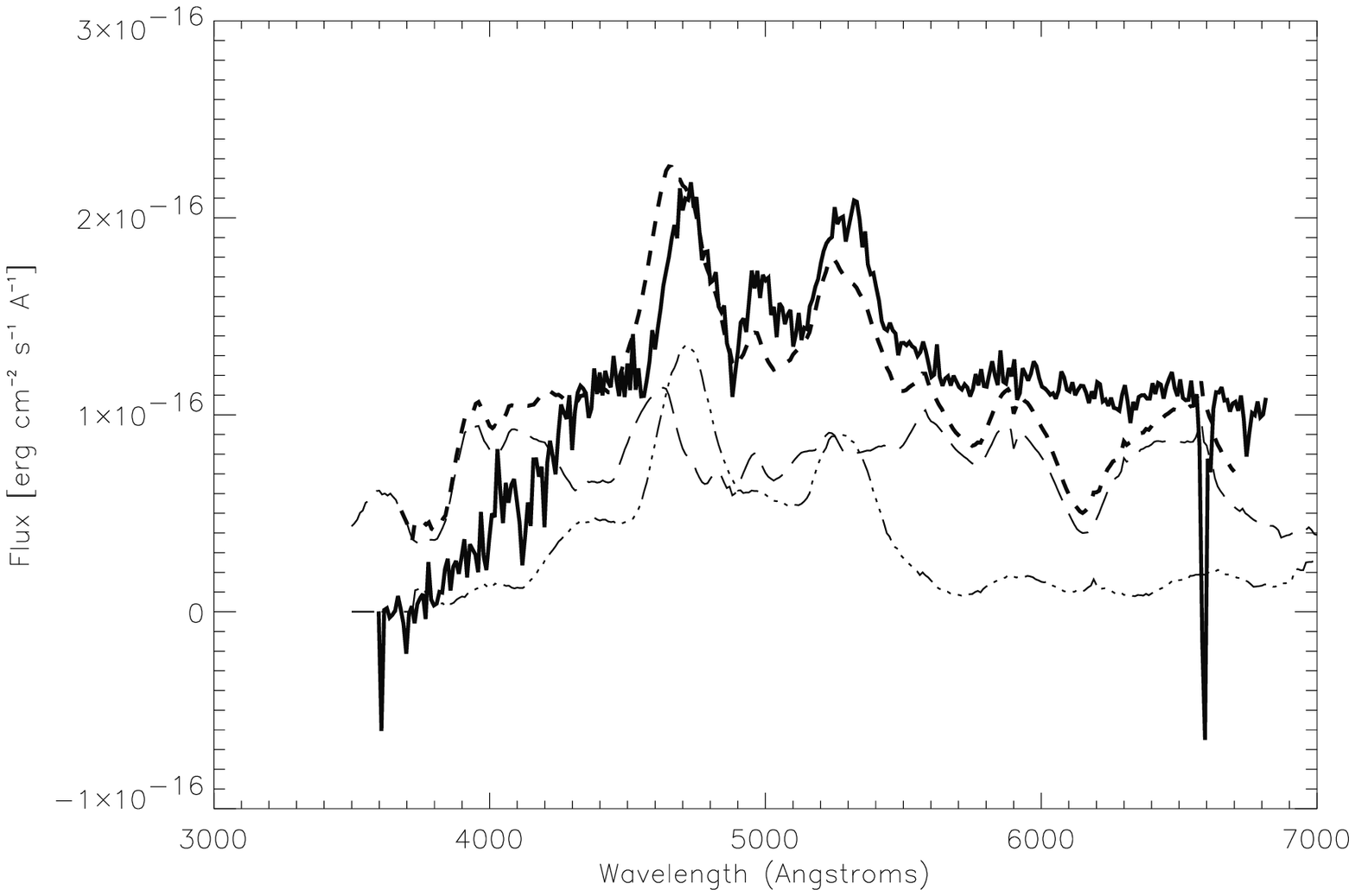}
\plotone{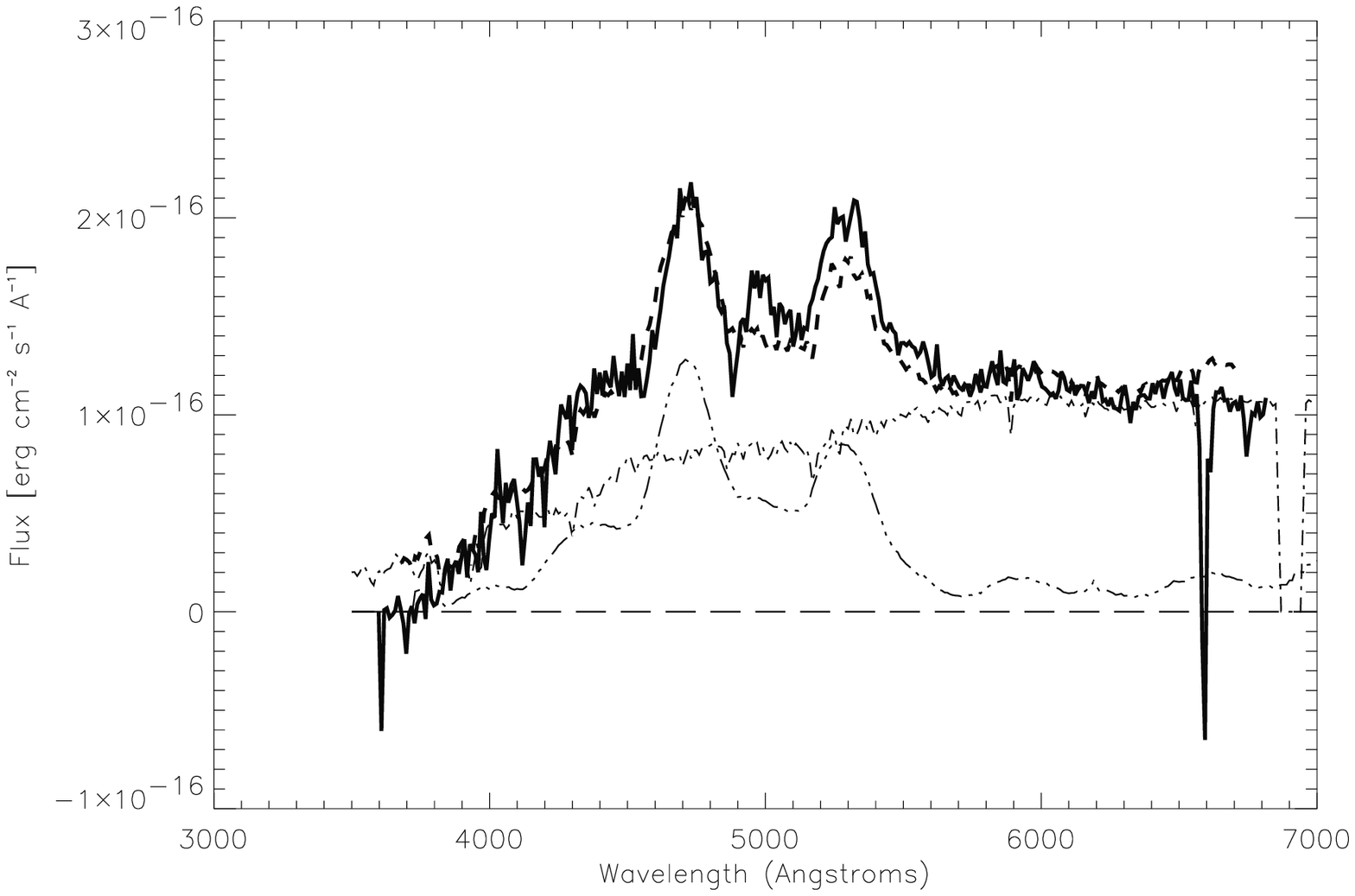}
\caption{The day 330 spectrum of SN 1989B. The spectrum is fit as in 
Figure 7. The nebular spectrum is a day 338 spectrum of SN 1996X, the 
stellar spectrum is of a G9-K0V type star. The spectrum created from the 
stellar spectrum fits the SN spectrum better than does the light echo  
spectrum. Reddening of E(B-V)=0.37 magnitude was assumed.}
\label{day330}
\end{figure}

\begin{figure}
\epsscale{0.90}
\plotone{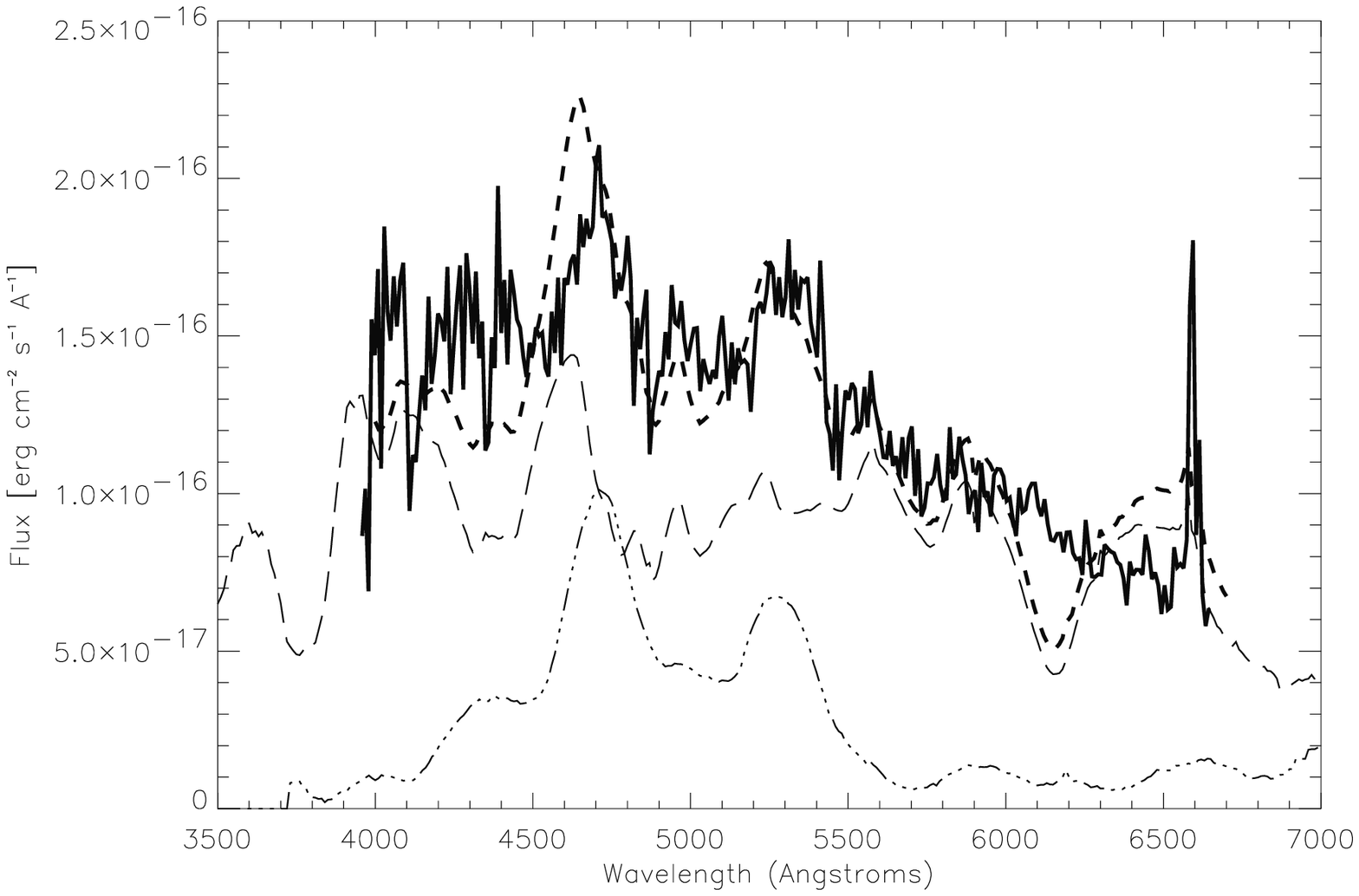}
\plotone{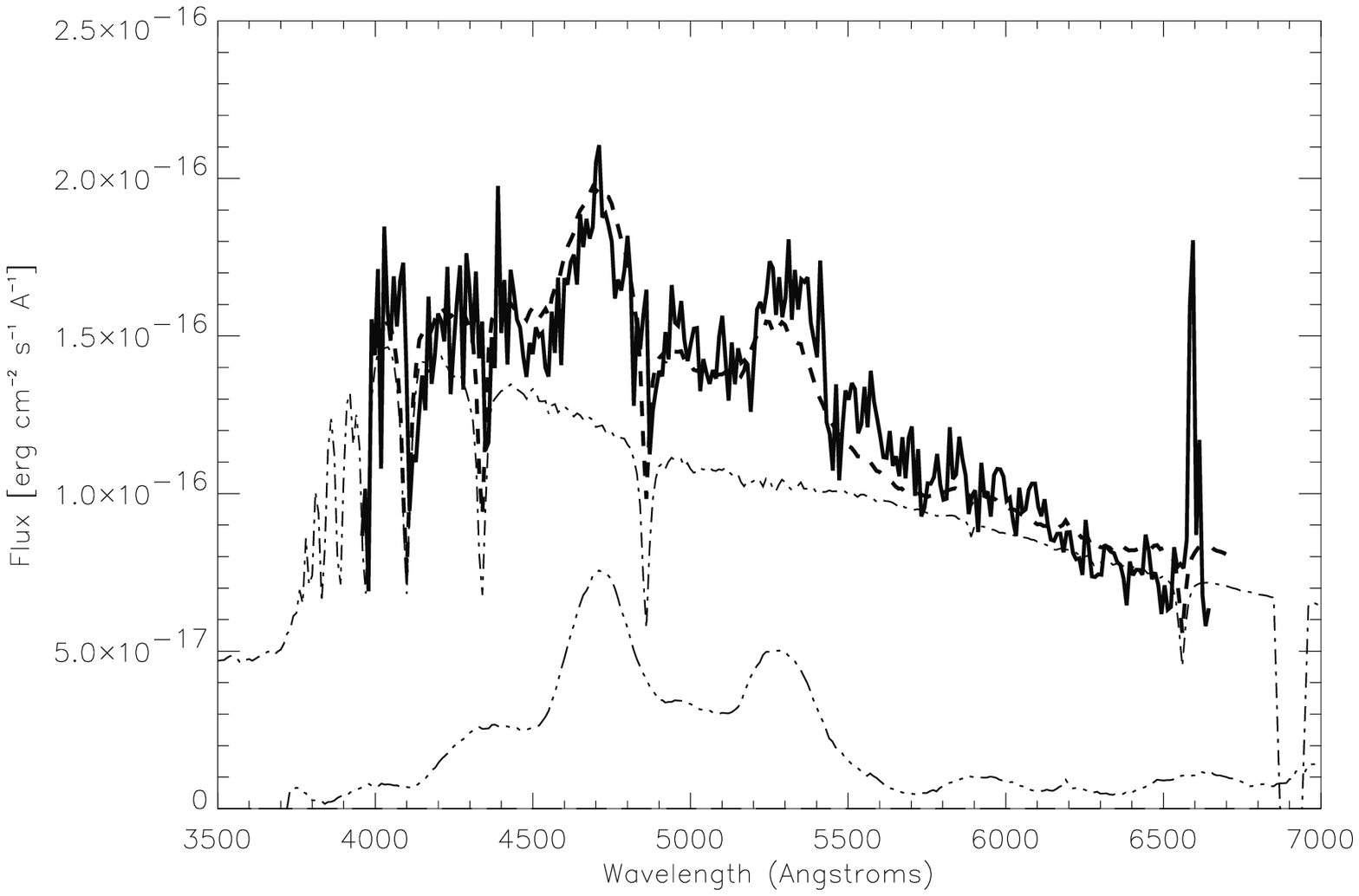}
\caption{The day 365 spectrum of SN 1989B. The spectrum is fit as in
Figure 7. The nebular spectrum is again a day 338 spectrum of SN 1996X, the 
stellar spectrum is of a A1-3V type star. The spectrum created from the 
stellar spectrum fits the SN spectrum better than does the light echo 
spectrum. Reddening of E(B-V)=0.37 magnitude was assumed for both fits.}
\label{day365}
\end{figure}


\begin{thebibliography}{}
\bibitem{barb90}Barbon, R., Benetti, S., Rosino, L., Cappellaro, E., Turatto, M. 
1990, {\it A\&A}, {\bf 237}, 79B.
\bibitem{boff99}Boffi, F.R., Sparks, W.B., Macchetto, F.D. 1999, {\it A\&A}, 
{\bf 138}, 253.
\bibitem{bolt89}Bolte, M., Saddlemyer, L., Mendes de Oliveira, C., Hodder, P. 
1989, {\it PASP}, {\bf 101}, 921.
\bibitem{bowe97}Bowers, E.J.C., Meikle, W.P.S., Geballe, T.R., Walton,
N.A., Pinto, P.A., Dhillon, V.S., Howell, S.B., \& Harrop-Allin, M.K.
1997, {\it MNRAS}, {\bf 290}, 663.
\bibitem{capp01}Cappellaro, E., Patat, F., Mazzali, P.A., Benetti, S., 
Danziger, J.I., Pastorello, A., Rizzi, L., Salvo, M., Turatto, M. 2001, {\it ApJ}, 
{\bf 549}, L215.
\bibitem{chev86}Chevalier, R.A. 1986, {\it ApJ}, {\bf 308}, 225.
\bibitem{cont00}Contardo, G., Leibudgut, B., Vacca, W.D. 2000, {\it A\&A},
{\bf 359}, 876.
\bibitem{garn01}Garnavich, P.A., et al. 2001, {\it BAAS}, {\bf 199}, 4701G.
\bibitem{holf98}H\H{o}flich, P., Wheeler, J. C.,  \& Thielemann, F.-K. 1998, 
{\it ApJ}, {\bf 495}, 617.
\bibitem{liu97}Liu, W., Jeffery, D.J., Schultz, D.R., Quinet, P., Shaw, J., 
Pindzola, M.S. 1997, {\it ApJ}, {\bf 489}, L141.
\bibitem{mtl99}Milne, P.A., The, L.-S., Leising, M.D. 1999, {\it ApJS}, {\bf 124}, 
503.
\bibitem{mtl01}Milne, P.A., The, L.-S., Leising, M.D. 2001, {\it ApJ}, {\bf 559}, 
1019.
\bibitem{ruiz93} Ruiz-Lapuente, P., Jeffery, D.J., Challis, P.M., Filippenko, A.V., 
Kirshner, R.P., Ho, L.C., Schmidt, B.P., Sanchez, F., Canal, R. 1993, {\it Nature}, {\bf 365}, 728. 
\bibitem{schm94}Schmidt, B.P., Kirshner, R.P., Leibundgut, B., Wells, L.A., 
Porter, A.C., Ruiz-Lapuente, P., Challis, P., Filippenko, A.V. 1994, {\it ApJ}, 
{\bf 434}, L19.
\bibitem{spar99}Sparks, W.B., Macchetto, F., Panagia, N., Boffi, F.R., Branch, D., 
Hazen, M.L., Della Valle, M. 1999, {\it ApJ}, {\bf 523}, 585.
\bibitem{well94}Wells, L.A., et al. 1994, {\it AJ}, {\bf 106}, 2233.
\end{thebibliography}
\end{document}